\documentclass{elsart}
\usepackage{natbib}
\usepackage{graphicx}

\begin{document}

\def\ms{M$_{\odot}$}
\def\zs{Z$_{\odot}$}
\def\mi{M$_{\rm IN}$}
\def\zi{Z}
\def\xi{Z$_{\rm IN}$}
\def\me{M$_{\rm ENV}$}
\def\nsn{N$_{\rm Ib,c}$/N$_{\rm II}$}
\def\mup{M$_{\rm Ib,c}$}
\def\lb{L$_{\rm B,\odot}$}
\def\be{$\beta^+$}
\def\ee{e$^+$}
\def\ne{N$_{e^+}$}
\def\co{$^{56}$Co }
\def\ti{$^{44}$Ti }

\runauthor{Nikos Prantzos}
\begin{frontmatter}
\title{On the 511 keV emission line of positron  annihilation in the Milky Way}

\author{Nikos Prantzos}

\address{Institut d'Astrophysique de Paris}

\begin{abstract}
I review our current understanding of positron sources in the Galaxy, 
on the basis of the reported properties of 
 the observed 511 keV annihilation line.
It is argued  here that most of the disk positrons propagate away fom the disk 
(due to the low density environment) and the 
resulting low surface brightness annihilation emission is currently 
undetectable by SPI/INTEGRAL. It is also argued that
a large fraction of the disk positrons
may be transported via the regular magnetic field of the Galaxy into the bulge and
annihilate there. These ideas may alleviate current difficulties in 
interepreting INTEGRAL results in a "conventional" framework.
\end{abstract}


\end{frontmatter}

\section{Introduction}

The origin of the Galactic electron-positron annihilation radiation
 remains problematic ever since the original detection of
its characteristic 511 keV line (e. g. Diehl et al. 2006 and references
therein). In particular, recent  observations  
of the line intensity and spatial morphology with the SPI instrument aboard
INTEGRAL put severe constraints on its origin, since
it appears that 1.5$\pm$0.1 10$^{43}$ \ee/s are annihilated in the bulge alone
and  0.3$\pm$0.2 10$^{43}$ \ee/s  in the disk, i.e. that
the bulge/disk   ratio of annihilating positrons is $B/D\sim$3-9 
(Kn\"odlseder et al. 2005).

In a recent work, Weidenspointer et al. (2008) find a significant
 asymmetry in the disk emission (factor 1.7 between positive and
 negative latitudes).\footnote{Note that this finding is not confirmed in a recent
 analysis by Bouchet et al. 2008. }They also find that the observed
 distribution of low-mass X-ray binaries in the hard state has a
 remarkable similarity to the 511 keV longitude profile; they
 suggest then that those objects may be the main sources of positrons in the
 disk, with a sizeable (but certainly insufficient) contribution to the bulge emission.

In this short review, I discuss the Galactic positron sources in  the
light of recent developments. I argue that,
{\it if} it is assumed  that {\it positrons annihilate near their
sources},  then none of the proposed positron production sites,
either conventional or ``exotic''ones, completely
satisfies the observational constraints. Furthermore,
I argue that a  large fraction of the disk positrons,
produced by thermonuclear supernovae (SNIa) or other sources,
may be transported via the regular magnetic field of the Galaxy into the bulge,
where they annihilate. This increases both the bulge positron 
annihilation rate and the
bulge/disk ratio, alleviating considerably  the constraints imposed by 
the  SPI/INTEGRAL data analysis. In fact, 
I argue that the SPI data are compatible with values of $B/D$ as low as 0.5, 
because positrons can propagate away from their sources and fill a rather 
large volume (of low surface brightness), 
much larger than the relatively thin disks adopted in 
the analysis of  SPI data. This property is crucial to the success 
of the scenario proposed here, which depends also on the poorly known
properties of the Galactic  magnetic field. In any case, the issue of
positron propagation in the ISM and the magnetic field of the Galaxy appears
to be crucial for our understanding of the 511 keV emission.

\section{Positron sources in the Galaxy}
Thermonuclear supernovae (SNIa) are prolific e$^+$ producers, releasing on average
3 10$^{54}$ positrons (from the decay of $\sim$0.7 \ms of $^{56}$Co), 
but  most of them annihilate inside the SN.
Assuming that  $N_{e^+}$ positrons escape each SNIa and that the rate of SNIa
in the Galactic bulge and disk {\it per unit stellar mass} is $ R_{Ia}$  
(given by observations of SNIa in external galaxies), one finds that
the \ee \ production rate from \co radioactivity of SNIa is 
$ L \ = \ M \ R_{Ia} \ N_{e^+} $,
where $M$ is the stellar mass of the system. 
$N_{e^+}$  is currently the major uncertainty of the problem. Original
estimates, based on late optical lightcurves of SNIa, gave  
$N_{e^+}\sim$8 10$^{52}$ or an escaping fraction $f\sim$0.03 (Milne et
al. 1999).\footnote{Observations of the late {\it bolometric}
lightcurves of two SNIa (Sollerman et al. 2004, Strinziger and Sollerman
2007) are compatible with zero escaping fraction (or
$N_{e^+} $=0) {\it in the framework of 1-D models for SNIa}. In 
"canonical'' 1-D models, $^{56}$Co is produced and remains in the
inner part of the SNIa. However, hydrodynamical 3-D models and recent
observations (Tanaka et al. 2008) suggest that a sizeable fraction of
$^{56}Co$ may be found in the outermost, highest velocity layers, from
which positrons may, perhaps, escape. In any case, the (presently
unknown) configuration of the SNIa magnetic field is crucial for the issue
of e$^+$ escape.} Taking into account the masses of the MW disk and
bulge, as well as the corresponding SNIa rates, this number leads to
a positron emissivity $L_D$=1.95$^{+0.98}_{-0.93}$ 10$^{43}$
\ee/s for the disk and $L_B$=0.17$^{+0.083}_{-0.081}$ 10$^{43}$
\ee/s for the bulge (see e.g. Prantzos 2006). 
Those estimates suggest that the disk \ee \
emissivity is  slightly larger than  the total galactic
\ee \ annihilation rate required by SPI observations, but it is much
larger than the one of the bulge, contrary to observations.

In the case of the disk positrons, a major source is undisputably the
radioactivity of $^{26}$Al. The observed decay of the $\sim$2 \ms \ of $^{26}$Al
per Myr (see Diehl, this volume) provides about 0.3 10$^{43}$ e$^+$/s,
i.e. close to the value given by the latest SPI data for the disk positron annihilation rate. 
However, the
1.8 MeV map of $^{26}$Al does not show the degree of asymmetry claimed in
Weidenspointner et al. (2008) for the disk 511 keV emission.

Among the other astrophysical sources of positrons, 
X-ray binaries (XRBs) or some related class of objects, appear as plausible
candidates. Low-mass XRBs (LMXRBs) 
were suggested in Prantzos (2004), who noticed
that their observed longitude distribution in the Galaxy is strongly
peaked towards the central regions, not unlike the one of the 511 keV emision. 
He also noticed, however, that most of
the strongest sources (counting for 80\% of the total Galactic X-ray flux)
are evenly distributed in the Galactic plane, with no preference for 
the bulge; if their positron emissivity scales with (some power of) their X-ray
flux, then LMXRBs cannot be at the origin of the bulge Galactic positrons.

According to Weidenspointner et al. (2008), the observed asymmetry in the disk 511 keV emission matches closely the Galactic distribution of LMXRBs in their hard state (factor 1.7 in both cases between positive and negative Galactic longitudes) and this similarity suggests that disk positrons mostly originate from this particular class of compacts objects. It should be noted, however, that in all probability, at least half of the disk 511 keV emissivity is due to $^{26}$Al, which displays only
a small  degree of asymmetry. The remaining 511 keV emission (i.e. once the $^{26}$Al contribution is removed) is certainly even more asymmetric (factor $>$2.5) and no known source matches such an asymmetric profile; thus, it seems premature to conclude that hard-state LMXRBs are at the origin of Galactic positrons.

Finally, Guessoum et al. (2006) find that another sub-class of XRBs, namely micro-quasars, are potentially interesting positron producers. Altough it is not yet known whether their jets are loaded mostly with positrons or protons, theoretical estimates evaluate their individual positron emissivity up to $\sim$10$^{41}$e$^+$/s. Coupled with their estimated number in the Milky Way (of the order of 100), that value suggests that micro-quasars may be interesting candidates.

If the Galactic SNIa positron emissivity evaluated in the first
paragraph of this section is close to the 
real one, but the source of the observed bulge emission turns out to
be  different, it should then  be a rather strange coincidence.
We argue here that transport of disk positrons to the bulge through the
Galactic magnetic field may inverse the Disk/Bulge ratio of \ee \ annihilation rates.
The arguments  are valid for any other source producing positrons of 
$\sim$1 MeV, such as those resulting from  radioactivity or  
X-ray binaries, or microquasar jets.

\section{Positron propagation in the Galaxy}

Positrons released in the ISM (especially in its less dense regions,
like e.g. outside spiral arms or outside the thin gaseous layer)
propagate affected by the Galactic magnetic field.
The large scale regular magnetic field (MF) of the Milky Way
 is composed of a toroidal (disk) component (probably
bisymmetric) and a poloidal (halo) component, probably in the form
of a A0 dipole (see Han 2004 and references therein also Fig. 1).
At Galactocentric distance r=8 kpc the toroidal component
has a strength of a few $\mu$G (Beck et al. 1996) and dominates
the poloidal one (a few tenths of  $\mu$G). However, the former varies 
as 1/r, while the latter as 1/r$^3$ and should therefore
dominate in the inner Galaxy.

Positron propagation is strongly affected by the
irregular (turbulent) component of the  galactic MF, which is comparable
in intensity with the regular one near the local disk. Unfortunately, the
properties of the irregular component away from the disk plane are even 
less well understood than those of the regular components (e.g. Han 2004).
Prouza and Smida (2003) assume that the 
turbulent component occupies 80\% of the volume inside spiral arms,
20\% of the volume outside spiral arms and within
vertical distance $|z|<$1.5 kpc and 
only 1\% of the volume at larger distances. 
On the other hand, cosmic ray propagation models
indicate that the size of the cosmic ray ``halo'' (CRH, i.e. the
region inside which cosmic rays diffuse on inhomogeneities of
the magnetic field) is $z_{CRH}>3$ kpc, based on
measurements of unstable/stable ratios of secondary nuclei
(Strong and Moskalenko 2001). If \ee \ escape from the CRH
then positron propagation at large distances from the disk will be dominated
by the regular MF, i.e. the poloidal field. In those conditions, a
fraction of the positrons  produced from disk SNIa will
ultimately find their way into the bulge.

Taking into account the SPI data (see Sec. 1) and the SNIa emissivity
of positrons in the disk and the bulge (Sec. 2) it turns out that
in order to explain the observed bulge annihilation rate
by SNIa, the fraction of disk positrons channeled to the bulge
must  be $f_{ESC}\sim$0.5, (i.e. $\sim$10$^{43}$ \ee/s 
from the disk have to join the 0.17 10$^{43}$ \ee/s 
produced in the bulge and annihilate in that region.

\begin{figure}
\centering
\includegraphics[width=0.5\textwidth]{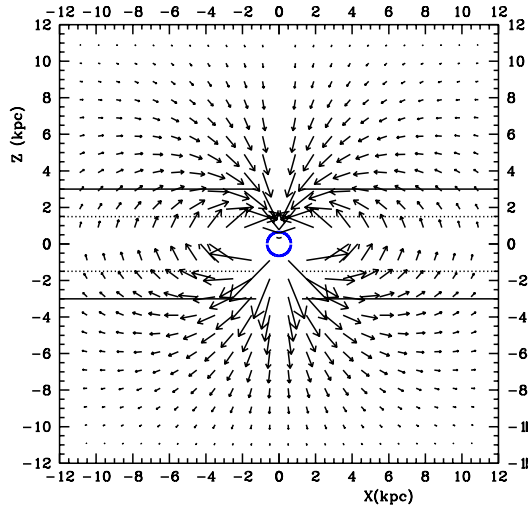}
\caption{\label{f3} A representation of the poloidal magnetic field of
the Milky Way in the x-z plane, {\it assuming} it is a A0 dipole. 
The central circle corresponds to the size of the 511 keV emitting
region (FWHM) seen by SPI aboard INTEGRAL. The irregular (turbulent)
magnetic field {\it is assumed} negligible outside the 
region between the horizontal lines ({\it dotted} 
lines are suggested in Prouza and Smida 2003, while {\it solid} lines are
lower limits to the cosmic ray halo size suggested in
Strong and Moskalenko 2001); positrons escaping from that region are 
presumably directed
by the dipole field lines towards the bulge.
}  
\end{figure}

This fraction may not be unreasonable. For instance, reacceleration of
\ee \ by shock waves of SNIa may considerably increase the
thermalization (travel) time of positrons $\tau_{SD}$.
On the other hand, the positron confinement time in the disk may be
shorter than the standard value of $\tau_{CONF}\sim$10$^7$ yr.
The reason is that, because of their low energy (and correspondingly low
gyroradius in the Galactic MF) 1 MeV positrons may diffuse very little on the 
density fluctuations of the MF, and thus they may escape more easily than 
the higher energy particles of standard Galactic cosmic rays. These arguments are
discussed further in Prantzos (2006), while  Jean et al. (2006) find
that low energy positrons may travel distances of several kpc in the hot, tenuous interstellar
medium (which  dominates away from the disk of the Milky Way).

In Prantzos (2006) it is also argued that the SNIa disk positrons may not only leave the disk,
but also enter the bulge by avoiding the ``magnetic mirror'' effect. The reason is that their
velocity has a dominant component which is always parallel to the  lines of the regular
magnetic field of the Galaxy.
 When positrons are still in the cosmic ray halo, they diffuse on the
 turbulent component of the MF at small scales, but at large scales
 their diffusive motion follows the regular (toroidal) component.
The configuration of the Galactic MF can vary only smoothly  between the regions
where the various components of the regular field (toroidal in the disk and poloidal
away from it) dominate. The toroidal
field changes smoothly into a poloidal one and positrons leaving the
former enter the latter with a velocity essentially parallel to its 
field lines;  this minimizes the losses due to the mirror effect.

\begin{figure}
\centering
\includegraphics[width=0.95\textwidth]{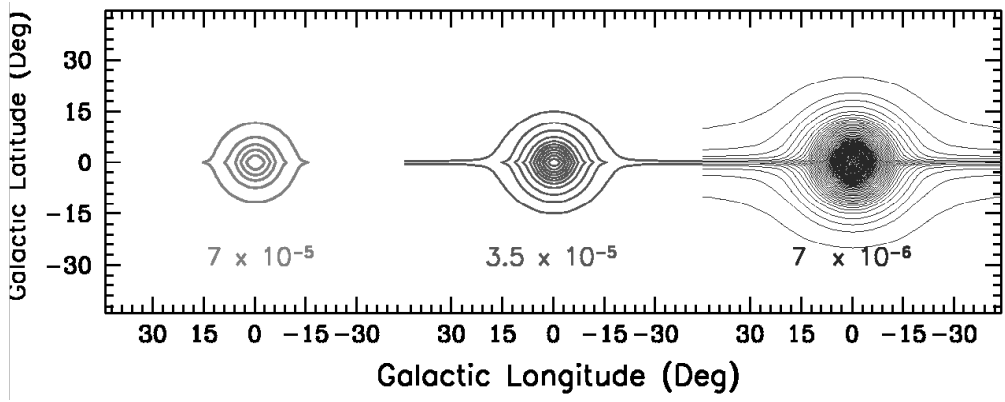}
\caption{\label{f4} Simulated profile of the 511 keV emission 
of the Milky Way in galactic coordinates, as seen at three different
 levels of detector sensisitivity (indicated in the bottom of each
 panel, in ph/cm$^2$/s). The adopted composite model for the positron 
annihilation (bulge + thin disk + thick disk) is described in Sec. 3.} 
\end{figure}

It is then concluded that 1) MeV positrons produced in the disk may propagate away from it,
and 2) they may also be chanelled to the bulge and annihilate there.
In Fig. 2 we calculate the Galactic gamma-ray profile from positron
annihilation, as seen with three different levels of sensitivity
(appearing in the bottom of each panel).
We assume that the Milky Way \ee \ annihilation rate results from i) a bulge with annihilation rate 
L=1.2 10$^{43}$ \ee/s (resulting from transfer of $\sim$50\% of the disk SNIa 
positrons plus those produced by the bulge SNIa population), ii) a
``thick disk'' (scaleheight 3 kpc) from the remaining SNIa positrons,
and iii) a thin disk (from positrons released by $^{26}$Al decay and
annihilating in the thin gaseous layer). Of course, such a symmetric
model does not reproduce the small disk asymmetry reported in Weidenspointer
et al. (2008).

We show quantitatively (and with more details in Prantzos 2006)
that the SPI/INTEGRAL data are fully compatible 
even with bulge/disk 
ratio of \ee \ annihilation rates 
lower than 1, provided that sufficiently (but not unreasonably) extended positron 
distributions are considered. We stress, in that respect, that positrons
can be treated similar to radioactive particles (since they have to slow down for a 
characteristic time $\tau_{SD}$ before annihilating), so that {\it 
the resulting 511 keV profile reflects essentially 
the distribution of propagated positrons} and not the product of their
density times the gas density.

The model proposed here exploits a range of possibilities, given our poor understanding of the Galactic
magnetic field and of the propagation of low energy positrons in it. However, its 
assumptions may  be tested,
through future observational and theoretical developments.
Systematic multi-wavelength studies of SNIa, including the infrared, will determine
 ultimately the typical positron yield of those objects. A small
511 keV emission outside the bulge is currently seen by SPI/INTEGRAL
 and, given enough exposure, the spatial extent 
of that emission will be determined (either by INTEGRAL or by a future instrument);
 an extended disk emission will prove that positrons travel indeed far away from their
 sources. Finally, the morphology of the Galactic magnetic field, and especially the 
presence of a poloidal component, will be put on more sound basis through further
 measurements (e.g. Han 2004).

\def\aj{AJ}
\def\apj{ApJ}
\def\apjs{ApJS}
\def\aap{A\&A}
\def\aaps{A\&AS}


\begin{thebibliography}{}


\bibitem {} Beck R., Brandenburg A., Moss D., et al. 1996, ARAA 34, 155

\bibitem {} Bouchet, L., Jourdain, E., Roques, J. P., 2008, ApJ 679, 1315


\bibitem {} Diehl R., Prantzos N., von Ballmoos P., 2006,  NuPhA 777, 70

\bibitem {} Guessoum, N., Jean, P., Prantzos, N., 2006, A\&A 457, 753 



\bibitem {} Han J. L., 2004, in ``The Magnetized Interstellar Medium'', 
             B. Uyaniker et al. (eds.), Copernicus GmbH 
             (Katlenburg-Lindau), p.3
             
\bibitem {} Jean P., Kn\"odlseder J., Gillard W., et al. 2006, \aap, 445, 579 


\bibitem {} Kn\"odlseder J., Jean P., Lonjou, V., et al. 2005,  \aap 441, 513




\bibitem {} Milne, P., The, L.-S., Leising, M., 1999, ApJS 124, 503


\bibitem {} Prantzos N., 2004, in ``The INTEGRAL Universe'', ESA SP-552, p. 15

\bibitem {} Prantzos N., 2006, A\&A, 449, 869

\bibitem {} Prouza M., Smida R., 2003, A\&A 410, 1


\bibitem {} Sollerman, J, Lindahl, J., Kozma, C., et al. 2004, A\&A 428, 555

\bibitem {} Strinziger, M., Sollerman, J., 2007, A\&A 407, L1


\bibitem {} Strong A., Moskalenko I., 2001, Adv. Sp. Res. 27, 717

\bibitem {} Tanaka, M., Mazzali, P., Benetti, S. et al. 2008, ApJ 677, 448

\bibitem {} Weidenspointner, G., Skinner, G., Jean, P., et al., 2008, Nature 451, 159

\end{thebibliography}
\end{document}